\documentclass[aps, showpacs, showkeys,nofootinbib,floatfix]{revtex4}

\usepackage{amssymb}
\usepackage{amsmath}
\usepackage{graphicx}

\begin{document}

\title{Cosmological constraints on generalized Chaplygin gas model: Markov Chain Monte Carlo
approach}

\author{Lixin Xu}
\email{lxxu@dlut.edu.cn}
\author{Jianbo Lu}
\email{lvjianbo819@163.com}

\affiliation{School of Physics and Optoelectronic Technology, Dalian
University of Technology, Dalian, 116024, P. R. China}

\begin{abstract}

We use the Markov Chain Monte Carlo method to  investigate a global
constraints on  the generalized Chaplygin gas (GCG) model as the
unification of dark matter and dark energy from  the latest
observational data: the Constitution dataset of type supernovae Ia
(SNIa), the observational Hubble data (OHD),  the cluster X-ray gas
mass fraction, the baryon acoustic oscillation (BAO), and the
 cosmic microwave background (CMB) data. In a
non-flat universe,  the constraint results for GCG model are,
 $\Omega_{b}h^{2}=0.0235^{+0.0021}_{-0.0018}$ ($1\sigma$) $^{+0.0028}_{-0.0022}$ $(2\sigma)$,
 $\Omega_{k}=0.0035^{+0.0172}_{-0.0182}$ ($1\sigma$) $^{+0.0226}_{-0.0204}$ $(2\sigma)$,
 $A_{s}=0.753^{+0.037}_{-0.035}$ ($1\sigma$) $^{+0.045}_{-0.044}$ $(2\sigma)$,
 $\alpha=0.043^{+0.102}_{-0.106}$ ($1\sigma$) $^{+0.134}_{-0.117}$ $(2\sigma)$,
 and $H_{0}=70.00^{+3.25}_{-2.92}$ ($1\sigma$) $^{+3.77}_{-3.67}$ $(2\sigma)$, which is more
stringent than the previous results for constraint on GCG model
parameters. Furthermore, according to the information criterion, it
seems that the current observations much support $\Lambda$CDM model
relative to the GCG model.
\end{abstract}


\maketitle

\section{$\text{Introduction}$}

 { Recently, mounting cosmic observations suggest
that the expansion of present universe is speeding up rather than
slowing down \cite{SNeCMBLSS}.  And they indicates that baryon
matter component is about 5\% for total energy density, and about
95\% energy density in universe is invisible. Considering the
four-dimensional  standard cosmology, this accelerated expansion for
universe predict that dark energy (DE) as an exotic component with
negative pressure is filled in universe. And it is shown that DE
takes up about two-thirds of the total energy density from cosmic
observations. On the other hand, in theory many kinds of DE models
have already been constructed in order to explore the DE properties.
For a review on DE models, please see Refs. \cite{DEmodels}.

It is well known that the generalized Chaplygin gas (GCG) model have
been widely studied for interpreting the accelerating universe
\cite{GCG}\cite{GCGpapers}. The most interesting property for this
scenario is that, two unknown dark sections in universe--dark energy
and dark matter  can be unified by using an exotic equation of
state. In this paper, we use the Markov Chain Monte Carlo (MCMC)
technique to constrain the GCG model from  the latest observational
data: the Constitution dataset \cite{397Constitution} including 397
type Ia supernovae (SNIa), the observational Hubble data (OHD)
\cite{OHD},  the cluster X-ray gas mass fraction
\cite{ref:07060033},  the measurement results of baryon acoustic
oscillation (BAO) from Sloan Digital Sky Survey (SDSS) and Two
Degree Field Galaxy Redshift Survey (2dFGRS)
\cite{SDSS}\cite{ref:Percival2}, and the current cosmic microwave
background (CMB) data from five-year WMAP \cite{5yWMAP}.

\section{$\text{generalized Chaplygin gas model}$}

 For GCG model, the energy density
$\rho$ and pressure $p$ are related by the equation of state
\cite{GCG}
\begin{equation}
p_{GCG}=-\frac{A}{\rho_{GCG}^{\alpha}},\label{e1}
\end{equation}
where $A$  and $\alpha$ are parameters in the model. By using the
energy momentum conservation equation in the
Friedmann-Robertson-Walker (FRW) cosmology,  $d(\rho
a^{3})=-pd(a^{3})$, the energy density of GCG fluid can be given
\begin{equation}
\rho_{GCG}=\rho _{0GCG}[A_s+(1-A_s)(1+z)^{3(1+\alpha )}]^{\frac
1{1+\alpha }},\label{e2}
\end{equation}
where $a$ is the scale factor, $z$ describes the redshift, and
A$_s=\frac A{\rho _0^{1+\alpha }}$. It is easy to see that $A_{s}$
denotes the current equation of state for GCG fluid
\cite{GCG-As}\cite{GCG-As-alpha}. Since the GCG fluids behaves as
dust at early stage and as dark energy at later stage, it can be
considered as a scenario of the unification of dark matter and dark
energy \cite{GCGunificationDMDE}. Considering a non-flat universe is
filled with three components: the GCG component, the baryon matter
component, and the radiation component, we have the total energy
density, $ \rho_{total}=\rho _{GCG}+\rho _{b}+\rho_{r}+\rho_{k}$.
Making use of the Friedmann equation, the Hubble parameter $H$ is
expressed as
\begin{equation}
H^2=\frac{8\pi G \rho_{total}}{3}=H_0^2\{(1-\Omega
_{b}-\Omega_{r}-\Omega_{k})[A_s+(1-A_s)(1+z)^{3(1+\alpha )}]^{\frac
1{1+\alpha }}+\Omega
_{b}(1+z)^{3}+\Omega_{r}(1+z)^{4}+\Omega_{k}(1+z)^{2}\},\label{e3}
\end{equation}
where  $H_{0}=100h$ km s$^{-1}$Mpc$^{-1}$ is the present Hubble
constant, $\Omega_{b}$, $\Omega_{r}$, and $\Omega_{k}$ denote
dimensionless baryon matter, radiation, and curvature density,
respectively.

\section{$\text{Current observational data and cosmological constraint}$}\label{constraint-method}
In this section, we introduce how the currently available data are
used to constrain the model in our calculation.

\subsection{Type Ia supernovae}

We constrain the parameters with  Constitution dataset
\cite{397Constitution} including 397 SNIa  \cite{397Constitution},
which is obtained by adding 90 SNIa from CfA3 sample to 307 SNIa
Union  sample \cite{307Union}. CfA3 sample are all from the
low-redshift SNIa, $z<0.08$, and these 90 SNIa are calculated with
using the same Union cuts. The addition of CFA3 sample increases the
number of nearby SNIa and reduces the statistical uncertainties. The
theoretical distance modulus $\mu(z)_{th}$ is defined as
\begin{equation}
\mu_{th}(z)=5\log_{10}[D_{L}(z)]+\mu_{0}.
\end{equation}
In this expression $D_{L}(z)$ is the Hubble-free luminosity distance
$H_0 d_L(z)/c$,
 and
\begin{eqnarray}
d_L(z)&=&\frac{c(1+z)}{\sqrt{|\Omega_k|}}sinn[\sqrt{|\Omega_k|}\int_0^z\frac{dz'}{H(z')}], \nonumber\\
\mu_0&\equiv&42.38-5\log_{10}h.\nonumber
\end{eqnarray}
where $sinn(\sqrt{|\Omega_k|}x)$ respectively denotes
$\sin(\sqrt{|\Omega_k|}x)$, $\sqrt{|\Omega_k|}x$,
$\sinh(\sqrt{|\Omega_k|}x)$ for $\Omega_k<0$, $\Omega_k=0$ and
$\Omega_k>0$.
 Additionally, the observed distance moduli $\mu_{obs}(z_i)$ of SNIa at $z_i$ is
\begin{equation}
\mu_{obs}(z_i) = m_{obs}(z_i)-M,
\end{equation}
where $M$ is their absolute magnitudes.

For the SNIa dataset, the best fit values of the parameters $p_s$
  can be determined by a likelihood analysis, based on
the calculation of
\begin{eqnarray}
\chi^2(p_s,M^{\prime})\equiv \sum_{SNIa}\frac{\left\{
\mu_{obs}(z_i)-\mu_{th}(p_s,z_i)\right\}^2} {\sigma_i^2} \ \ \ \ \ \ \ \ \ \  \ \ \ \ \ \ \nonumber\\
=\sum_{SNIa}\frac{\left\{ 5 \log_{10}[D_L(p_s,z_i)] - m_{obs}(z_i) +
M^{\prime} \right\}^2} {\sigma_i^2}, \ \ \ \ \label{eq:chi2}
\end{eqnarray}
where $p_s$ denotes the parameters contained in the model,
$M^{\prime}\equiv\mu_0+M$ is a nuisance parameter which includes the
absolute magnitude and the parameter $h$. The nuisance
  parameter $M^{\prime}$ can be marginalized over
analytically \cite{ref:SNchi2} as
\begin{equation}
\bar{\chi}^2(p_s) = -2 \ln \int_{-\infty}^{+\infty}\exp \left[
-\frac{1}{2} \chi^2(p_s,M^{\prime}) \right] dM^{\prime},\nonumber
\label{eq:chi2marg}
\end{equation}
resulting to
\begin{equation}
\bar{\chi}^2 =  A - \frac{B^2}{C} + \ln \left(
\frac{C}{2\pi}\right), \label{eq:chi2mar}
\end{equation}
with
\begin{eqnarray}
&&A=\sum_{SNIa} \frac {\left\{5\log_{10}
[D_L(p_s,z_i)]-m_{obs}(z_i)\right\}^2}{\sigma_i^2},\nonumber\\
&& B=\sum_{SNIa} \frac {5
\log_{10}[D_L(p_s,z_i)]-m_{obs}(z_i)}{\sigma_i^2},\nonumber
\\
&& C=\sum_{SNIa} \frac {1}{\sigma_i^2}\nonumber.
\end{eqnarray}
Relation (\ref{eq:chi2}) has a minimum at the nuisance parameter
value $M^{\prime}=B/C$, which contains information of the values of
$h$ and $M$. Therefore, one can extract the values of $h$ and $M$
provided the knowledge of one of them. Finally, note that the
expression
\begin{equation}
\chi^2_{SNIa}(p_s,B/C)=A-(B^2/C),\label{eq:chi2SN}
\end{equation}
which coincides to (\ref{eq:chi2mar}) up to a constant, is often
used in the likelihood analysis \cite{ref:SNchi2}, and thus in this
case the results will not be affected by a flat $M^{\prime}$
distribution.

\subsection{Observational Hubble data}

The observational Hubble data are based on differential ages of the
galaxies \cite{ref:JL2002}. In \cite{ref:JVS2003}, Jimenez {\it et
al.} obtained an independent estimate for the Hubble parameter using
the method developed in \cite{ref:JL2002}, and used it to constrain
the EOS of dark energy. The Hubble parameter depending on the
differential ages as a function of redshift $z$ can be written in
the form of
\begin{equation}
H(z)=-\frac{1}{1+z}\frac{dz}{dt}.
\end{equation}
So, once $dz/dt$ is known, $H(z)$ is obtained directly \cite{OHD}.
By using the differential ages of passively-evolving galaxies from
the Gemini Deep Deep Survey (GDDS) \cite{ref:GDDS} and archival data
\cite{ref:archive1,ref:archive2,ref:archive3,ref:archive4,ref:archive5,ref:archive6},
Simon {\it et al.} obtained $H(z)$ in the range of $0\lesssim z
\lesssim 1.8$ \cite{OHD}. The twelve observational Hubble data from
\cite{ref:0905,ref:0907} are list in Table \ref{Hubbledata}.
\begin{table}[htbp]
\begin{center}
\begin{tabular}{c|llllllllllll}
\hline\hline
 $z$ &\ 0 & 0.1 & 0.17 & 0.27 & 0.4 & 0.48 & 0.88 & 0.9 & 1.30 & 1.43 & 1.53 & 1.75  \\ \hline
 $H(z)\ ({\rm km~s^{-1}\,Mpc^{-1})}$ &\ 74.2 & 69 & 83 & 77 & 95 & 97 & 90 & 117 & 168 & 177 & 140 & 202  \\ \hline
 $1 \sigma$ uncertainty &\ $\pm 3.6$ & $\pm 12$ & $\pm 8$ & $\pm 14$ & $\pm 17$ & $\pm 60$ & $\pm 40$
 & $\pm 23$ & $\pm 17$ & $\pm 18$ & $\pm 14$ & $\pm 40$ \\
\hline
\end{tabular}
\end{center}
\caption{\label{Hubbledata} The observational $H(z)$
data~\cite{ref:0905,ref:0907}.}
\end{table}
In addition, in \cite{ref:0807}, the authors took the BAO scale as a
standard ruler in the radial direction, obtain three more additional
data: $H(z=0.24)=79.69\pm2.32, H(z=0.34)=83.8\pm2.96,$ and
$H(z=0.43)=86.45\pm3.27$.

 The best fit values of the model parameters from
observational Hubble data  are determined by minimizing
\cite{chi2hub}
\begin{equation}
\chi_{Hub}^2(p_s)=\sum_{i=1}^{15} \frac{[H_{th}(p_s;z_i)-H_{
obs}(z_i)]^2}{\sigma^2(z_i)},\label{eq:chi2H}
\end{equation}
where  $H_{th}$ is the predicted value for the Hubble parameter,
$H_{obs}$ is the observed value, $\sigma(z_i)$ is the standard
deviation measurement uncertainty, and the summation is over the
$15$ observational Hubble data points at redshifts $z_i$.

\subsection{The X-ray gas mass fraction constraints}
According to the X-ray cluster gas mass fraction observation, the
baryon mass fraction in clusters of galaxies (CBF) can be utilized
to constrain cosmological parameters. The X-ray gas mass fraction,
$f_{gas}$, is defined as the ratio of the X-ray gas mass to the
total mass of a cluster, which is a constant and independent on the
redshift. In the framework of the $\Lambda CDM$ reference cosmology,
the X-ray gas mass fraction is presented as \cite{ref:07060033}
\begin{eqnarray}
&&f_{gas}^{\Lambda CDM}(z)=\frac{K A \gamma
b(z)}{1+s(z)}\left(\frac{\Omega_b}{\Omega_m}\right)
\left[\frac{D_A^{\Lambda CDM}(z)}{D_A(z)}\right]^{1.5},\ \ \ \
\label{eq:fLCDM}
\end{eqnarray}
where $A$ is the angular correction factor, which is caused by the
change in angle for the current test model $\theta_{2500}$ in
comparison with that of the reference cosmology
$\theta_{2500}^{\Lambda CDM}$:
\begin{eqnarray}
&&A=\left(\frac{\theta_{2500}^{\Lambda
CDM}}{\theta_{2500}}\right)^\eta \approx
\left(\frac{H(z)D_A(z)}{[H(z)D_A(z)]^{\Lambda CDM}}\right)^\eta,
\end{eqnarray}
here, the index $\eta$ is the slope of the $f_{gas}(r/r_{2500})$
data within the radius $r_{2500}$, with the best-fit average value
$\eta=0.214\pm0.022$ \cite{ref:07060033}. And the proper (not
comoving) angular diameter distance is given by
\begin{eqnarray}
&&D_A(z)=\frac{c}{(1+z)\sqrt{|\Omega_k|}}\mathrm{sinn}[\sqrt{|\Omega_k|}\int_0^z\frac{dz'}{H(z')}].
\end{eqnarray}
It is clear that this quantity is related with $d_{L}(z)$ by
\begin{equation}
D_A(z)=\frac{d_{L}(z)}{(1+z)^2}.\nonumber
\end{equation}
 For GCG model, since it is considered as the
unification of dark matter and dark energy, we do not have dark
matter in this model. So, the matter density is not explicitly
included in the background equation (\ref{e3}).
 Following the Ref. \cite{EffMatterGCG}, we use an relation between $\Omega_{m}$ and $A_{s}$:
$\Omega_{m}=1-\Omega_{r}-\Omega_{k}-A_{s}(1-\Omega_{b}-\Omega_{r}-\Omega_{k})$.
This parameter, $\Omega_{m}$, is an estimate of the "matter"
component of the GCG fluid with the baryon density. And for the BAO
and CMB constraint methods in the following, we also take this
expression of $\Omega_{m}$.

In equation (\ref{eq:fLCDM}), the parameter $\gamma$ denotes
permissible departures from the assumption of hydrostatic
equilibrium, due to non-thermal pressure support; the bias factor
$b(z)= b_0(1+\alpha_b z)$ accounts for uncertainties in the cluster
depletion factor; $s(z)=s_0(1 +\alpha_s z)$ accounts for
uncertainties of the baryonic mass fraction in stars and a Gaussian
prior for $s_0$ is employed, with $s_0=(0.16\pm0.05)h_{70}^{0.5}$
\cite{ref:07060033}; the factor $K$ is used to describe the combined
effects of the residual uncertainties, such as the instrumental
calibration and certain X-ray modelling issues, and a Gaussian prior
for the 'calibration' factor is considered by $K=1.0\pm0.1$
\cite{ref:07060033};

Following the method in Ref. \cite{ref:CBFchi21,ref:07060033} and
adopting the updated 42 observational $f_{gas}$ data in Ref.
\cite{ref:07060033}, the best fit values of the model parameters for
the X-ray gas mass fraction analysis are determined by minimizing,
\begin{eqnarray}
&&\chi^2_{CBF}=\sum_i^N\frac{[f_{gas}^{\Lambda
CDM}(z_i)-f_{gas}(z_i)]^2}{\sigma_{f_{gas}}^2(z_i)}.
\end{eqnarray}

\subsection{Baryon acoustic oscillation}

The baryon acoustic oscillations are detected in the clustering of
the combined 2dFGRS and SDSS main galaxy samples, and measure the
distance-redshift relation at $z = 0.2$. Additionally, baryon
acoustic oscillations in the clustering of the SDSS luminous red
galaxies measure the distance-redshift relation at $z = 0.35$. The
observed scale of the BAO calculated from these samples, as well as
from the combined sample, are jointly analyzed using estimates of
the correlated errors to constrain the form of the distance measure
$D_V(z)$ \cite{ref:Okumura2007,ref:Percival2}

\begin{equation}
D_V(z)=\left[(1+z)^2 D^2_A(z) \frac{cz}{H(z)}\right]^{1/3}.
\label{eq:DV}
\end{equation}
The peak positions of the BAO depend on the ratio of $D_V(z)$ to the
sound horizon size at the drag epoch (where baryons were released
from photons) $z_d$, which can be obtained by using a fitting
formula \cite{27Eisenstein}:
\begin{eqnarray}
&&z_d=\frac{1291(\Omega_mh^2)^{-0.419}}{1+0.659(\Omega_mh^2)^{0.828}}[1+b_1(\Omega_bh^2)^{b_2}],
\end{eqnarray}
with
\begin{eqnarray}
&&b_1=0.313(\Omega_mh^2)^{-0.419}[1+0.607(\Omega_mh^2)^{0.674}], \\
&&b_2=0.238(\Omega_mh^2)^{0.223}.
\end{eqnarray}
In this paper, we use the data of $r_s(z_d)/D_V(z)$ extracted from
the Sloan Digitial Sky Survey (SDSS) and the Two Degree Field Galaxy
Redshift Survey (2dFGRS) \cite{ref:Okumura2007}, which are listed in
Table \ref{baodata}, where $r_s(z)$ is the comoving sound horizon
size
\begin{eqnarray}
r_s(z)&&{=}c\int_0^t\frac{c_sdt}{a}=c\int_0^a\frac{c_sda}{a^2H}=c\int_z^\infty
dz\frac{c_s}{H(z)} \nonumber\\
&&{=}\frac{c}{\sqrt{3}}\int_{0}^{1/(1+z)}\frac{da}{a^2H(a)\sqrt{1+(3\Omega_b/(4\Omega_\gamma)a)}},
\end{eqnarray}
where $c_s$ is the sound speed of the photon$-$baryon fluid
\cite{ref:Hu1, ref:Hu2}:
\begin{eqnarray}
&&c_s^{-2}=3+\frac{4}{3}\times\frac{\rho_b(z)}{\rho_\gamma(z)}=3+\frac{4}{3}\times(\frac{\Omega_b}{\Omega_\gamma})a,
\end{eqnarray}
and here $\Omega_\gamma=2.469\times10^{-5}h^{-2}$ for
$T_{CMB}=2.725K$.

\begin{table}[htbp]
\begin{center}
\begin{tabular}{c|l}
\hline\hline
 $z$ &\ $r_s(z_d)/D_V(z)$  \\ \hline
 $0.2$ &\ $0.1905\pm0.0061$  \\ \hline
 $0.35$  &\ $0.1097\pm0.0036$  \\
\hline
\end{tabular}
\end{center}
\caption{\label{baodata} The observational $r_s(z_d)/D_V(z)$
data~\cite{ref:Percival2}.}
\end{table}
Using the data of BAO in Table \ref{baodata} and the inverse
covariance matrix $V^{-1}$ in \cite{ref:Percival2}:
\begin{eqnarray}
&&V^{-1}= \left(
\begin{array}{cc}
 30124.1 & -17226.9 \\
 -17226.9 & 86976.6
\end{array}
\right),
\end{eqnarray}
thus, the $\chi^2_{BAO}(p_s)$ is given as
\begin{equation}
\chi^2_{BAO}(p_s)=X^tV^{-1}X,\label{eq:chi2BAO}
\end{equation}
where $X$ is a column vector formed from the values of theory minus
the corresponding observational data, with
\begin{eqnarray}
&&X= \left(
\begin{array}{c}
 \frac{r_s(z_d)}{D_V(0.2)}-0.190533 \\
 \frac{r_s(z_d)}{D_V(0.35)}-0.109715
\end{array}
\right),
\end{eqnarray}
and $X^t$ denotes its transpose.

\subsection{Cosmic microwave background}

The CMB shift parameter $R$ is provided by \cite{ref:Bond1997}
\begin{equation}
R(z_{\ast})=\sqrt{\Omega_m H^2_0}(1+z_{\ast})D_A(z_{\ast})/c,
\end{equation}
which is related to the second distance ratio
$D_A(z_\ast)H(z_\ast)/c$ by a factor $\sqrt{1+z_{\ast}}$. The
redshift $z_{\ast}$ (the decoupling epoch of photons) is obtained
using the fitting function \cite{Hu:1995uz}
\begin{equation}
z_{\ast}=1048\left[1+0.00124(\Omega_bh^2)^{-0.738}\right]\left[1+g_1(\Omega_m
h^2)^{g_2}\right],\nonumber
\end{equation}
where the functions $g_1$ and $g_2$ read
\begin{eqnarray}
g_1&=&0.0783(\Omega_bh^2)^{-0.238}\left(1+ 39.5(\Omega_bh^2)^{0.763}\right)^{-1},\nonumber \\
g_2&=&0.560\left(1+ 21.1(\Omega_bh^2)^{1.81}\right)^{-1}.\nonumber
\end{eqnarray}
In additional, the acoustic scale is related to the first distance
ratio, $D_A(z_\ast)/r_s(z_\ast)$ (One can define an  angular scale
of the sound horizon at decoupling epoch), and is defined as
\begin{eqnarray}
&&l_A\equiv(1+z_{\ast})\frac{\pi D_A(z_{\ast})}{r_s(z_{\ast})},
\end{eqnarray}
where  a factor of $1+z_{\ast}$ arises because $D_A(z)$ is the
proper (physical) angular diameter distance, whereas
$r_{s}(z_{\ast})$ is the comoving sound horizon at $z_{\ast}$. Using
the data of $l_A, R, z_\ast$ in \cite{5yWMAP} and their covariance
matrix of $[l_A(z_\ast), R(z_\ast), z_\ast]$, we can calculate the
likelihood $L$ as $\chi^2_{CMB}=-2\ln L$:
\begin{eqnarray}
&&\chi^2_{CMB}=\bigtriangleup d_i[Cov^{-1}(d_i,d_j)[\bigtriangleup
d_i]^t],
\end{eqnarray}
where $\bigtriangleup d_i=d_i-d_i^{data}$ is a row vector, and
$d_i=(l_A, R, z_\ast)$.\\

\section{$\text{Method and results}$}

According to the descriptions in  section \ref{constraint-method},
for our calculations the total likelihood function is written as
$L\propto e^{-\chi^2/2}$, here the $\chi^2$ equals
\begin{eqnarray}
\chi^2=\chi^2_{SNIa}+\chi^2_{OHD}+\chi^2_{CBF}+\chi^2_{BAO}+\chi^2_{CMB}.
\end{eqnarray}
In our analysis, we perform a global fitting on determining the
cosmological parameters using a  MCMC. The MCMC code is listed in
the publicly available {\bf CosmoMC} package \cite{ref:MCMC} written
in Fortran 90. In addition, the likelihood of $f_{gas}$ has been
included in the modified CosmoMC. For the analysis of  X-ray cluster
gas mass fraction, we seek help from the online Fortran 90 code
\cite{ref:0409574,ref:07060033,ref:modifiedMCMC}, and correct the
patch for $f_{gas}$ after some nontrivial crosschecks. For each MCMC
calculation on GCG model, we run 8 independent chains comprising of
50000-60000 chain elements. The average acceptance rate is about
35\%. To get the converged results, we test the convergence of the
chains by typically getting $R - 1$ to be less than 0.03.

\begin{table}
\begin{center}
\begin{tabular}{cc   cc  cc  }
 \hline\hline parameters & flat GCG                              & non-flat GCG&   flat $\Lambda$CDM &   non-flat $\Lambda$CDM& \\ \hline
 $\Omega_{b}h^2$  &  $0.0233^{+0.0023 +0.0029}_{-0.0016 -0.0020}$   & $0.0235^{+0.0021 +0.0028}_{-0.0018 -0.0022}$ &      $-$&     $-$&\\
 $\Omega_{k} $    &  $-$                                           & $0.0035^{+0.0172 +0.0226}_{-0.0182 -0.0204}$  &      $-$                                 & $0.0002^{+0.0083 +0.0146}_{-0.0119 -0.0127}$& \\
$\Omega_{m} $    &  $-$                                           & $-$                                            &  $0.280^{+0.029 +0.039}_{-0.030 -0.038}$ & $0.278^{+0.027 +0.037}_{-0.031 -0.039}$&  \\
 $A_{s}$          &  $0.760^{+0.029 +0.034}_{-0.039 -0.046}$       & $0.753^{+0.037 +0.045}_{-0.035 -0.044}$ &        $-$&     $-$&\\
 $\alpha$         &  $0.033^{+0.066 +0.096}_{-0.071 -0.087}$       & $0.043^{+0.102 +0.134}_{-0.106 -0.117}$ &       $-$&     $-$& \\
 $H_0$            &  $69.97^{+2.87 +3.48}_{-2.78 -3.08}$           & $70.00^{+3.25 +3.77}_{-2.92 -3.67}$    &     $70.01^{+2.51 +3.19}_{-2.01 -2.69}$&     $70.24^{+2.55 +3.36}_{-3.34 -3.90}$& \\  \hline
 $\chi^2_{min}(\chi^{2}_{min}/dof)$    & 519.342 (1.139)            &  519.371 (1.144)&       520.351 (1.139)        &  520.302 (1.141)&       \\
\hline\hline
\end{tabular}
\caption{The data fitting results of GCG and $\Lambda$CDM model
parameters with $1\sigma$ and $2\sigma$ confidence levels for flat
and non-flat universe.}\label{tab:resultsGCG}
\end{center}
\end{table}

\begin{figure}[!htbp]
  \includegraphics[width=9cm]{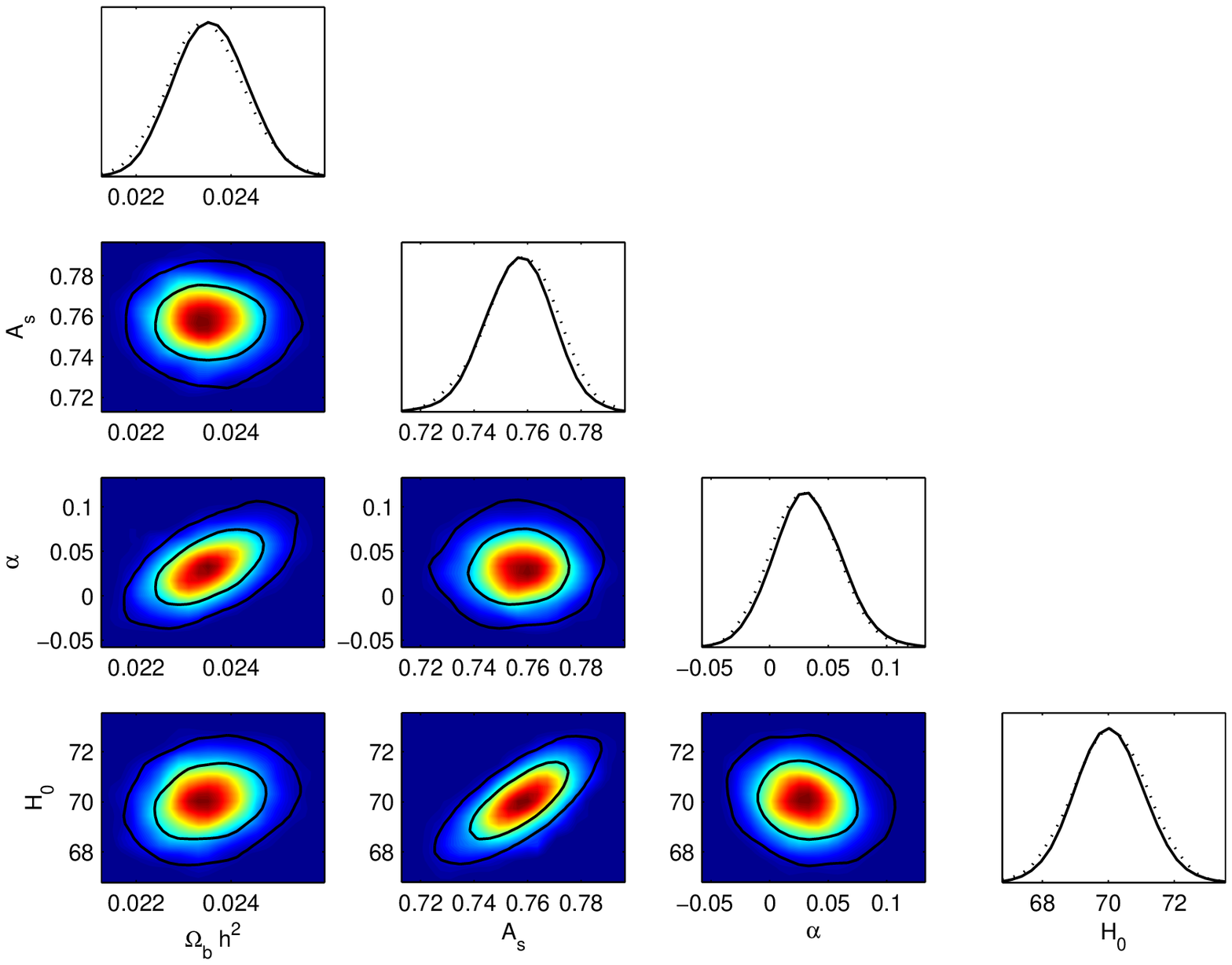}~
  \includegraphics[width=11cm]{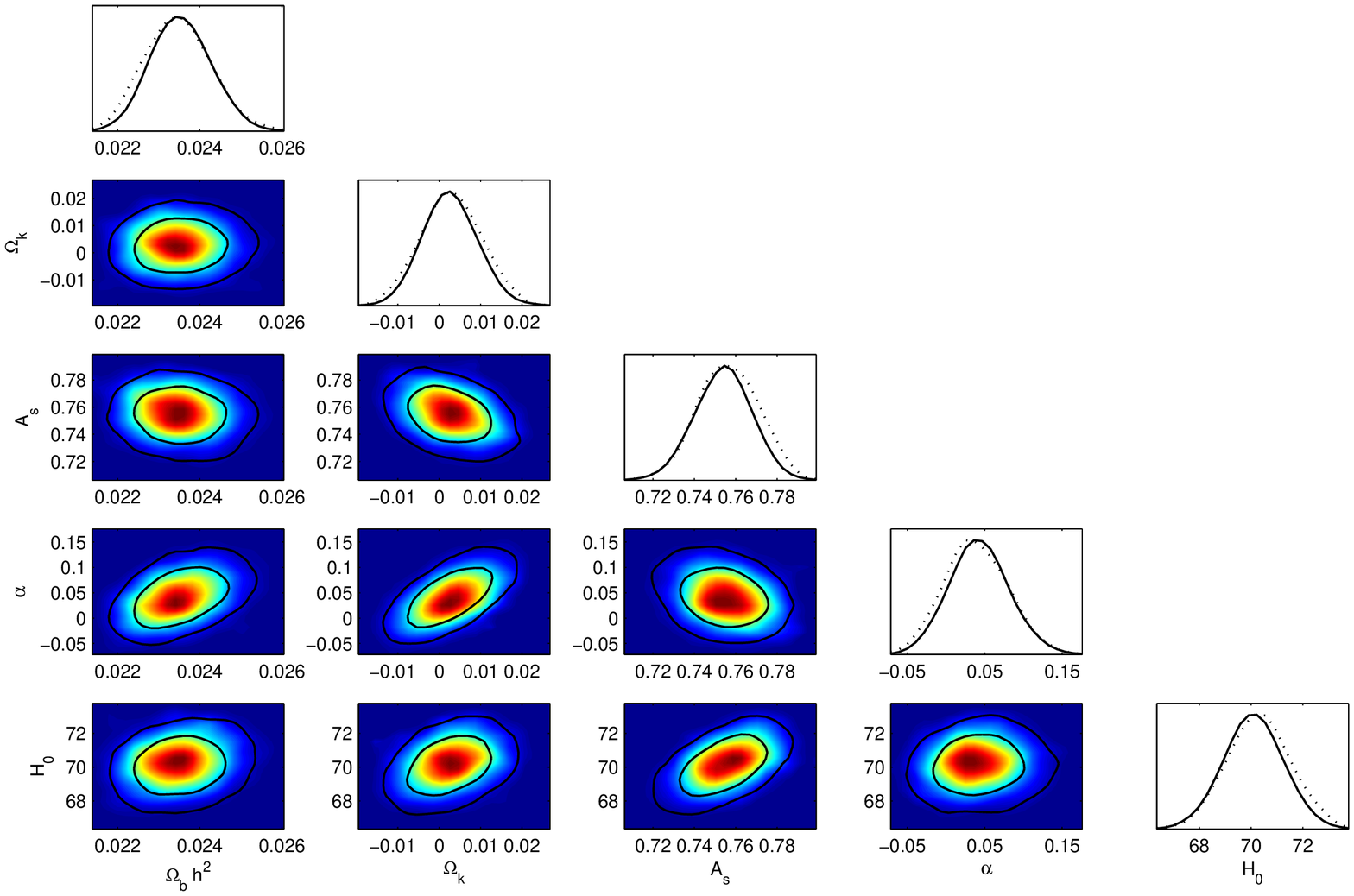}\\
 \caption{1-D constraints on model parameters and 2-D contours on these parameters with $1\sigma, 2\sigma$
  confidence levels between each other in  flat (left) and non-flat (right)  GCG model.}\label{fig:contoursGCG}
\end{figure}

\begin{figure}[!htbp]
  \includegraphics[width=5.5cm]{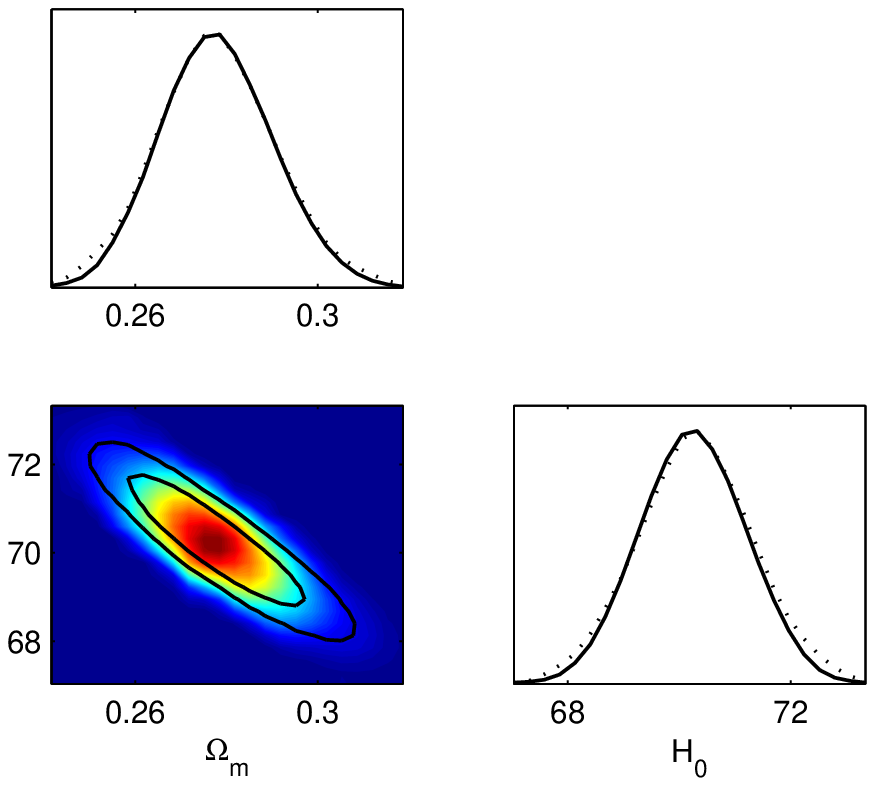}~~~~~~~~~~~~~~~~~~~~~~
  \includegraphics[width=6cm]{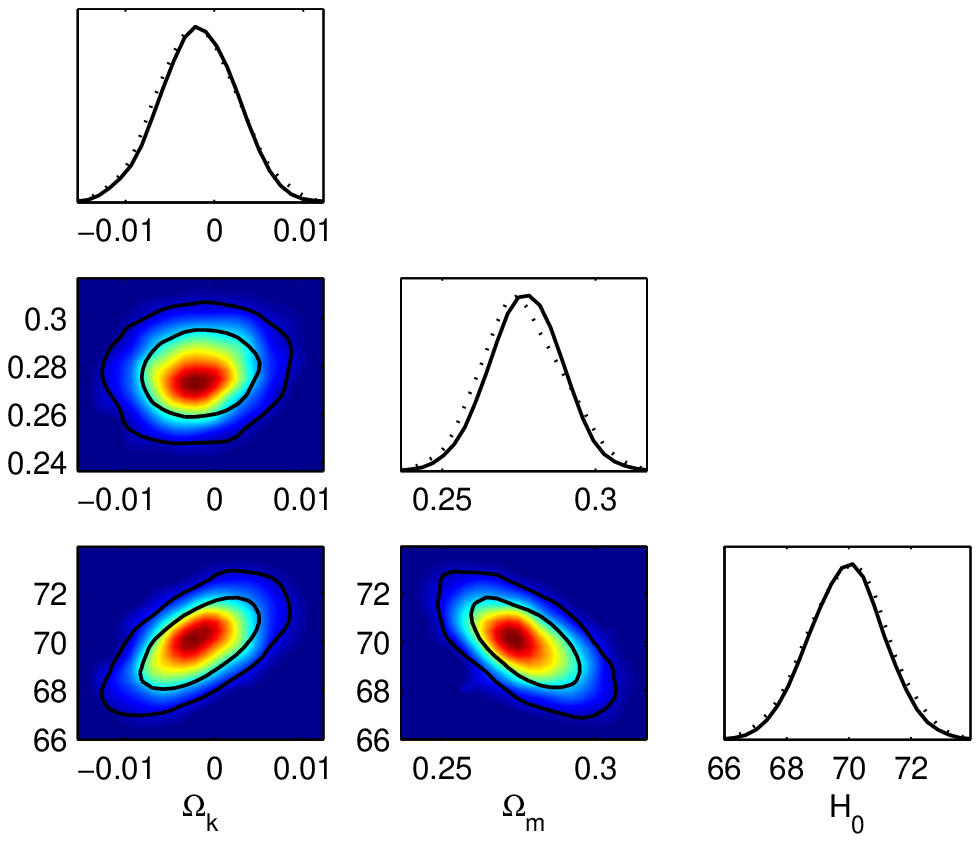}\\
 \caption{1-D constraints on model parameters and 2-D contours on these parameters with $1\sigma, 2\sigma$
  confidence levels between each other in  flat (left) and non-flat (right)  $\Lambda$CDM model.}\label{fig:contoursLCDM}
\end{figure}

In Fig. \ref{fig:contoursGCG}, we show one dimensional probability
distribution of each parameter and two dimensional plots for
parameters between each other for the GCG model in flat and non-flat
universe. Dotted lines are mean likelihoods of samples, and solid
lines are marginalized probabilities for $1D$ distribution in the
figure. According to Fig. \ref{fig:contoursGCG}, the constraint
results on the best fit values of cosmological parameters with
$1\sigma$ and $2\sigma$ confidence levels  are listed in Table
\ref{tab:resultsGCG}. From this table, we can see that this
constraint on GCG model parameter,
$\Omega_{b}h^2=0.0233^{+0.0023}_{-0.0016}$ ($1\sigma$)
$^{+0.0029}_{-0.0020}$ $(2\sigma)$, $A_{s}=0.760^{+0.029}_{-0.039}$
($1\sigma$) $^{+0.034}_{-0.046}$ $(2\sigma)$ and
 $\alpha=0.033^{+0.066}_{-0.071}$ ($1\sigma$) $^{+0.096}_{-0.087}$
 $(2\sigma)$ for a flat universe is more stringent than the result in Refs.
 \cite{GCG-As-alpha}\cite{constraintGCG}.
 Furthermore, we also consider the constraint on GCG model in a non-flat
 universe, and the constraint result are: $\Omega_{b}h^2=0.0235^{+0.0021}_{-0.0018}$ ($1\sigma$)
$^{+0.0028}_{-0.0022}$ $(2\sigma)$,
$\Omega_{k}=0.0035^{+0.0172}_{-0.0182}$ ($1\sigma$)
$^{+0.0226}_{-0.0204}$ $(2\sigma)$,
 $A_{s}=0.753^{+0.037}_{-0.035}$ ($1\sigma$) $^{+0.045}_{-0.044}$ $(2\sigma)$,
 and  $\alpha=0.043^{+0.102}_{-0.106}$ ($1\sigma$)
 $^{+0.134}_{-0.117}$ $(2\sigma)$. From our results above, it is clear that the best fit parameter $\alpha$
 has a very small value, and its value is a little bigger than zero.
 Furthermore, for the case of $\alpha=0$, it is included in the $1\sigma$ confidence
 level around the best fit parameter. One knows for $\alpha=0$, the GCG model reduces to the cosmic concordance
 model, $\Lambda$CDM. Thus, it seems that the current cosmic observations indicate the concordance model should be favored when
we apply the present observed data to constrain a more complex GCG
model. Then in the following, we also consider the  constraint on
$\Lambda$CDM model  from the above datasets, and the results are:
 $\Omega_{m}=0.280^{+0.029}_{-0.030}$ ($1\sigma$) $^{+0.039}_{-0.038}$ $(2\sigma)$,
 $H_{0}=70.01^{+2.51}_{-2.01}$ ($1\sigma$) $^{+3.19}_{-2.69}$
 $(2\sigma)$ with $\chi^2_{min}=520.351$ for a flat universe,
  and  $\Omega_{k}=0.0002^{+0.0083}_{-0.0119}$ ($1\sigma$) $^{+0.0146}_{-0.0127}$ $(2\sigma)$,
 $\Omega_{m}=0.278^{+0.027}_{-0.031}$ ($1\sigma$) $^{+0.037}_{-0.039}$
 $(2\sigma)$, $H_{0}=70.24^{+2.55}_{-3.34}$ ($1\sigma$) $^{+3.36}_{-3.90}$
 $(2\sigma)$ with $\chi^2_{min}=520.302$ for a non-flat case.  It can be seen that though $\Omega_{k}$ is not exactly equal to zero,
  the curvature density parameter has
 a very  small value, and the best fit result shows
 a close universe.

We also calculate the values of $\chi_{min}^2/dof$  for GCG and
$\Lambda$CDM model in Table \ref{tab:resultsGCG},  where the value
of dof (degree of freedom) equals to the number of observational
data points minus the number of parameters.  From Table
\ref{tab:resultsGCG}, we can see that for using the large data
points, the difference of $\chi^{2}_{min}/dof$ for the different
models is not obvious. It appears that  for applying the quantity
$\chi_{min}^2/dof$ to rate the goodness of models, it is not a good
(or refined) method. So, in the following we also use the objective
information criteria (IC) to estimate the quality of the models. The
Akaike information criteria (AIC) and Bayesian information criteria
(BIC) are respectively defined as
\begin{equation}
AIC=-2\ln {\cal L}_{max}+2K,\label{AIC}
\end{equation}
\begin{equation}
BIC=-2\ln {\cal L}_{max}+K\ln n,\label{BIC}
\end{equation}
where ${\cal L}_{max}$ is the highest likelihood in the model with
the best fit parameters, $K$ is the number of estimable parameters,
$n$ is the number of data points in the fit\footnote{For small
sample sizes, there is a corrected version of the AIC,
AIC$_{c}$=AIC$+2K(K-1)/(n-K-1)$ \cite{correctAIC}, which is
important for $n/K\lesssim 40$. Obviously, in our case, this
correction is negligible.}. The term, $-2\ln {\cal
L}_{max}=\chi^{2}_{min}$, measures the quality of model fit, while
the terms including $K$ interpret model complexity. The absolute
value of the criterion for a single model has no meaning, and only
the relative values between different models are interesting.
Considering several candidate models, the one that minimizes the AIC
(or BIC) is usually considered the best. Comparing with the best
one, the difference in AIC (or BIC) for other one model is expressed
as $\Delta$AIC $ = \Delta \chi_{min}^{2} + 2\Delta K$ (or
$\Delta$BIC = $\Delta \chi_{min}^{2} + \Delta K \ln n$). Thus one
can assess the strength
 of the models. The rules for judging the AIC model
selection are \cite{AICselection}: when $0\leq$
$\Delta$AIC$_{i}$$\leq 2$
 model $i$ has almost the same support from the data
as the best model, for $2\leq $ $\Delta$AIC$_{i}$$\leq 4$, model $i$
is supported considerably less, and with $\Delta$AIC$_{i}$$
>10$
 model $i$ is practically irrelevant. For BIC, one has:
 a $\Delta$BIC
 of more than 2 (or 6) relative to the best one
is considered $^{"}$unsupported$^{"}$  (or $^{"}$strongly
unsupported$^{"}$) from observational data \cite{BICselection}.

For $\Lambda$CDM and GCG model in the  flat and non-flat universe,
the IC values against the model are listed in table \ref{AICBIC}.
According to the table, AIC shows that the flat and non-flat GCG
model
  are supported considerably less by current observational
 data. For BIC selection method, it seems that  a more complex GCG
model is not necessary to explain the current data. In addition, we
can see that the current observations much support a flat-geometry
universe. According to Ref. \cite{AICBICcompare}, one knows that in
the limit of large data points  ($\ln n > 2$), AIC tends to favor
models with more parameters while BIC tends to penalize them. Here
it should be noticed that for the used data in our analysis, $\ln n
=6.129$, so for BIC the punishment on much-parameter models is more
stern.

 \begin{table}
 \vspace*{-12pt}
 \begin{center}
 \begin{tabular}{c c  c c c} \hline
  Case model                     &  Free model parameters                                    & $\chi^{2}_{min}$~~~   & $\Delta AIC$ ~~~ & $\Delta BIC$     \\ \hline
  flat $\Lambda$CDM              &  $\Omega_{m}$, $H_{0}$                                     & 520.351         ~~~   & 0           ~~~  &  0               \\
  non-flat $\Lambda$CDM          &  $\Omega_{k}$, $\Omega_{m}$, $H_{0}$                       & 520.302         ~~~   &1.951        ~~~  &  6.080           \\
  flat GCG                       &  $\Omega_{b}$, $A_{s}$, $\alpha$,  $H_{0}$                 & 519.342         ~~~   &2.991        ~~~  &  11.249           \\
  non-flat GCG                   &  $\Omega_{k}$, $\Omega_{b}$, $A_{s}$, $\alpha$, $H_{0}$    & 519.371         ~~~   &5.020        ~~~  &  17.407           \\ \hline
 \end{tabular}
 \end{center}
 \caption{Information criteria results } \label{AICBIC}
 \end{table}

\section{$\text{Conclusion}$}

The constraints on the flat and non-flat GCG model as the
 unification of dark matter and dark energy are studied in this paper by using
  the latest observational data: the  Constitution dataset
including 397 SNIa, the Hubble parameter data, the cluster X-ray gas
mass fraction, the baryon acoustic oscillation and  the five-year
WMAP data. The constraint  on GCG model parameters are more
 stringent than the previous papers
 \cite{GCG-As-alpha}\cite{constraintGCG}. According to the
 constraint results, since the best fit values of parameters $\alpha$ and $\Omega_{k}$ are near to
 zero, it seems that the current observations tends to
 make the GCG model reduce to  the flat
 $\Lambda$CDM model. Furthermore, according to the IC, we can get
 the same result. In addition, we also make a stringent
 constraint on $\Lambda$CDM model, and in a flat  $\Lambda$CDM model  it is shown that the cosmic age
 is about, $t_{age}(Gyr)=13.725^{+0.099}_{-0.141}$ ($1\sigma$) $^{+0.134}_{-0.165}$
 $(2\sigma)$, for using a tophat prior as 10 Gyr $< t_{age} <$ 20 Gyr in our calculation. At last,  for the  CBF constraint method, as a reference
we list the best-fit values of the parameters in
  $f_{gas}$: $K=0.9919, \eta=0.2089, \gamma=1.0299, b_0=0.7728,
 \alpha_b=-0.0582, s_0=0.1656, \alpha_s=0.1128$ for the flat universe and
 $K=0.9871, \eta=0.2114, \gamma=1.0507, b_0=0.7749, \alpha_b=-0.0950,
 s_0=0.1741, \alpha_s=0.0194$ in the non-flat case.

\section*{Acknowledgments}
The data fitting is based on the  publicly available {\bf CosmoMC}
package including a  MCMC code. This work is supported by the
National Natural Science Foundation of China (Grant No.10703001),
and Specialized Research Fund for the Doctoral Program of Higher
Education (Grant No. 20070141034).


\begin{thebibliography}{*}
 \bibitem{SNeCMBLSS} A.G. Riess {\it et al}, 1998 {\it Astron. J.} {\bf 116} 1009 [arXiv:astro-ph/9805201];
  S. Perlmutter {\it et al}, 1999 {\it Astrophys. J.} {\bf  517} 565;
  D.N. Spergel {\it et al},  2003 {\it Astrophys. J. Suppl.} {\bf 148} 175 [arXiv:astro-ph/0302209];
  A.C. Pope {\it et al},  2004 {\it Astrophys. J.}  {\bf 607} 655 [arXiv:astro-ph/0401249].

\bibitem{DEmodels}
 S. Weinberg, 1989 {\it Mod. Phys. Rev.} {\bf 61} 527;
  B. Ratra and P.J.E. Peebels, 1988 {\it Phys. Rev. D.} {\bf 37} 3406;
   R.R. Caldwell, M. Kamionkowski and N. N. Weinberg, 2003 {\it Phys. Rev. Lett.} {\bf91} 071301 [arXiv:astro-ph/0302506];
    B. Feng, X.L. Wang and X.M. Zhang, 2005  {\it Phys. Lett. B} {\bf 607} 35 [arXiv:astro-ph/0404224];
     M. Li, 2004 {\it Phys. Lett. B}  {\bf 603} 1 [arXiv:hep-th/0403127];
      L.X. Xu, J.B. Lu and W.B. Li, 2009 {\it Eur. phys. J. C} {\bf64} 89;
       L.X. Xu, W.B. Li and  J.B. Lu, 2009 {\it Mod. phys. Lett. A} {\bf24} 1355.



\bibitem{GCG}
 A.Y. Kamenshchik, U. Moschella and V. Pasquier, 2001 {\it Phys. Lett. B} {\bf 511} 265 [arXiv:gr-qc/0103004];
  M.C. Bento, O. Bertolami and A.A. Sen, Phys. Rev. D 66 (2002) 043507 [arXiv:gr-qc/0202064].

\bibitem{GCGpapers}
 T. Barreiro, O. Bertolami and P. Torres,  2008 {\it Phys. Rev. D} \textbf{78} 043530 [arXiv:astro-ph/0805.0731];
 M. Makler, S.Q. Oliveira and I. Waga, 2003 {\it Phys. Lett. B} \textbf{555} 1;
 R. Bean, O. Dore, 2003 {\it Phys. Rev. D} \textbf{68}  023515;
 L. Amendola, L.F. Finelli, C. Burigana, D. Carturan, 2003  {\it J. Cosmol. Astropart. Phys.} \textbf{0307} 005;
 A. Dev, D. Jain, J.S. Alcaniz, 2004 {\it Astron. Astrophys.} \textbf{417} 847;
 M. Makler, S.Q. Oliveira and I. Waga, 2003 {\it Phys. Rev. D} {\bf 68} 123521;
 J.B. Lu {\it et al},  2008 {\it Phys. Lett. B}  {\bf 662}, 87;
 J.A.S. Lima, J.V. Cunha and J.S. Alcaniz, [arXiv:astro-ph/0611007].



\bibitem{397Constitution}
  M.~Hicken {\it et al.},  Astrophys. J. {\bf 700} 1097 (2009), [arXiv:astro-ph/0901.4804].


\bibitem{OHD}
 J. Simon, L. Verde and R. Jimenez, Phys. Rev. D {\bf 71} 123001 (2005) [astro-ph/0412269].


\bibitem{ref:07060033}
 S. W. Allen, D. A. Rapetti, R. W. Schmidt, H. Ebeling, R. G. Morris and A. C. Fabian, Mon. Not. Roy. Astron. Soc. {\bf 383} 879 (2008).

\bibitem{SDSS}
 D.J. Eisenstein {\it et al},  2005 {\it Astrophys. J.} {\bf 633}, 560 [arXiv:astro-ph/0501171].


\bibitem{ref:Percival2} W.J. Percival {\it et al.}, arXiv:0907.1660 [astro-ph.CO].


\bibitem{5yWMAP}
 E.~Komatsu {\it et al.},  Astrophys. J. Suppl. {\bf 180} 330 (2009) [arXiv:astro-ph/0803.0547];
 J. Dunkley,  et al., [arXiv:astro-ph/0803.0586].

\bibitem{GCG-As}
  L. Amendola, F. Finelli, C. Burigana, D. Carturan, JCAP 0307 (2003)  005,  arXiv:astro-ph/0304325.

\bibitem{GCG-As-alpha}
  P.X Wu, H.W. Yu, JCAP 0703 (2007) 015, arXiv:astro-ph/0701446.


\bibitem{GCGunificationDMDE}
 P.T. Silva, O. Bertolami, 2003  {\it Astrophys. J.} \textbf{599} 829, arXiv:astro-ph/0303353.

\bibitem{307Union}
 D. Rubin  {\it et al}, [arXiv:astro-ph/0807.1108].


\bibitem{ref:SNchi2}
 S. Nesseris  and L. Perivolaropoulos,  Phys. Rev. D 72 (2005) 123519 [astro-ph/0511040];
 L. Perivolaropoulos, Phys. Rev. D {\bf 71} 063503 (2005);
 E. Di Pietro and J. F. Claeskens, Mon. Not. Roy. Astron. Soc. {\bf 341} 1299 (2003);
 A.~C.~C.~Guimaraes, J.~V.~Cunha and J.~A.~S.~Lima, JCAP {\bf 0910} 010 (2009);
 J.B. Lu,  {\it Phys. Lett. B}  {\bf 680}, 404 (2009);
 M. Szydlowski and W. Godlowski, 2006 {\it Phys. Lett. B} {\bf633} 427  [arXiv:astro-ph/0509415];
 S. Nesseris and L. Perivolaropoulos,  2007 {\it JCAP} {\bf0702} 025, [arXiv:astro-ph/0612653];
 L. Perivolaropoulos,  2005 {\it Phys. Rev. D} {\bf71} 063503;
 E. Di Pietro and J. F. Claeskens, 2003 {\it Mon. Not. Roy. Astron. Soc.}  {\bf 341} 1299, [arXiv:astro-ph/0207332];
 U. Alam and V. Sahni, 2006  {\it Phys.Rev.D} \textbf{73} 084024;  L. Xu, W. Li and J. Lu, JCAP {\bf 0907} 031 (2009).


\bibitem{ref:JL2002}
 R. Jimenez and A. Loeb, Astrophys. J. {\bf 573} 37 (2002) [astro-ph/0106145].

\bibitem{ref:JVS2003}
 R. Jimenez, L. Verde, T. Treu and D. Stern, Astrophys. J. {\bf 593} 622 (2003) [astro-ph/0302560].


\bibitem{ref:GDDS}
 R. G. Abraham {\it et al.}, Astron. J. {\bf 127} 2455 (2004) [astro-ph/0402436].

\bibitem{ref:archive1}
 T. Treu, M. Stiavelli, S. Casertano, P. Moller and G. Bertin, Mon. Not. Roy. Astron. Soc. {\bf 308} 1037 (1999);
 R.G. Abraham  {\it et al},  2003  {\it Astron. J.} {\bf593} 622.

\bibitem{ref:archive2}
 T. Treu, M. Stiavelli, P. Moller, S. Casertano and G. Bertin, Mon. Not. Roy. Astron. Soc. {\bf 326} 221 (2001) [astro-ph/0104177].

\bibitem{ref:archive3}
 T. Treu, M. Stiavelli, S. Casertano, P. Moller and G. Bertin, Astrophys. J. Lett. {\bf 564} L13 (2002).

\bibitem{ref:archive4}
 J. Dunlop, J. Peacock, H. Spinrad, A. Dey, R. Jimenez, D. Stern and R. Windhorst, Nature {\bf 381} 581 (1996).

\bibitem{ref:archive5}
 H. Spinrad, A. Dey, D. Stern, J. Dunlop, J. Peacock, R. Jimenez and R. Windhorst, Astrophys. J. {\bf 484} 581 (1997).

\bibitem{ref:archive6}
 L. A. Nolan, J. S. Dunlop, R. Jimenez and A. F. Heavens, Mon. Not. Roy. Astron. Soc. {\bf 341} 464 (2003) [astro-ph/0103450].



\bibitem{ref:0905}
  A. G. Riess {\it et al.}, arXiv:0905.0695 [astro-ph.CO].

\bibitem{ref:0907}
  D. Stern {\it et al.}, arXiv:0907.3149 [astro-ph.CO].


\bibitem{ref:0807}
  E. Gaztanaga {\it et al.}, arXiv:0807.3551 [astro-ph.CO].


\bibitem{chi2hub}
 R. Lazkoz and E. Majerotto, 2007  {\it JCAP} {\bf0707} 015 [arXiv:astro-ph/0704.2606];
 J.B Lu, L.X Xu, M.L Liu and Y.X Gui, 2008 {\it Eur. Phys. J. C} {\bf58} 311 [arXiv:astro-ph/0812.3209];
 L. Samushia and B. Ratra, 2006 {\it Astrophys. J.} {\bf650} L5  [astro-ph/0607301];
 R. Jimenez, L. Verde, T. Treu and D. Stern, 2003 {\it Astrophys. J.} {\bf593} 622 [astro-ph/0302560].



\bibitem{EffMatterGCG}
 T. Barreiro, O. Bertolami, P. Torres, Phys.Rev.D 78, 043530, 2008 arXiv:0805.0731;
 M.C. Bento, O. Bertolami, A.A. Sen, Phys.Lett. B 575 (2003) 172-180 arXiv:astro-ph/0303538.



\bibitem{ref:CBFchi21}
 S. Nesseris and L. Perivolaropoulos, JCAP {\bf 0701} 018 (2007) [astro-ph/0610092].



\bibitem{ref:Okumura2007}
 D. J. Eisenstein {\it et al.}, Astrophys. J. {\bf 633} 560 (2005);
 W. J. Percival {\it et al.}, Mon. Not. Roy. Astron. Soc. {\bf 381} 1053 (2007).

\bibitem{27Eisenstein}
 D.J. Eisenstein and W. Hu, 1998 {\it Astrophys. J.} {\bf 496} 605 [arXiv:astro-ph/9709112]


\bibitem{ref:Hu1}
 W. Hu and N. Sugiyama, Astrophys. J. {\bf 444} 489 (1995) [arXiv:astro-ph/9407093].

\bibitem{ref:Hu2}
 W. Hu, M. Fukugita, M. Zaldarriaga and M. Tegmark, Astrophys. J. {\bf 549} 669 (2001) [arXiv:astro-ph/0006436].

\bibitem{ref:Bond1997}
 J.~R.~Bond, G.~Efstathiou and M.~Tegmark, Mon. Not. Roy. Astron. Soc.  {\bf 291} L33 (1997).

\bibitem{Hu:1995uz}
 W.~Hu and N.~Sugiyama, Astrophys. J. {\bf 471} 542 (1996).



\bibitem{ref:MCMC}
 A. Lewis and S. Bridle, Phys. Rev. D {\bf 66} 103511 (2002);
 URL: http://cosmologist.info/cosmomc/.


\bibitem{ref:0409574}
 D. Rapetti, S. W. Allen and J. Weller, Mon. Not. Roy. Astron. Soc. {\bf 360} 555 (2005).

\bibitem{ref:modifiedMCMC}
 URL: http://www.stanford.edu/~drapetti/fgas-module/

\bibitem{constraintGCG}
 Z.H. Zhu, 2004 {\it Astron. Astrophys.} {\bf 423}  421;
 P.X. Wu and H.W. Yu  2007 {\it Phys. Lett. B} {\bf 644} 16;
 J.B. Lu, Y.X. Gui and L.X. Xu, {\it Eur. phys. J. C} {\bf63} 349.

\bibitem{correctAIC}
 A. R. Liddle,   Mon. Not. Roy. Astron. Soc. Lett. 377, L74 (2007) [arXiv:astroph /0701113].

\bibitem{AICselection}
  M. Biesiada, J. Cosmol. Astron. Phys.  {\bf0702} 003 (2007),astro-ph/0701721


\bibitem{BICselection}
  A.R. Liddle,  Mon. Not. R. Astron. Soc. {\bf351} L49 (2004), astro-ph/0401198

\bibitem{AICBICcompare}
 D. Parkinson, S. Tsujikawa, B.A. Bassett, L. Amendola, Phys. Rev. D {\bf71}, 063524, (2005)


\end{thebibliography}
\end{document}